\documentclass[useAMS,usenatbib]{mn2e}

\usepackage{graphicx}

\newcommand{\degrees}{\ensuremath{^\circ}}

\def\QBH{${Q}_{B/H}$} 
\def\Z{${Q}_{B/H}^{(1)}$} 
\def\ZZ{${Q}^{(2)}_{B/H}$} 
\def\ZZZ{${Q}^{(3)}_{B/H}$} 
\def\ZZZZ{${Q}^{(4)}_{B/H}$} 
\def\Zi{${Q}^{(i)}_{B/H}$} 

\title[Comparison of bar strengths in optical and near-infrared]{Comparison of bar strengths in optical and near-infrared for the {\sc osubsgs} sample}

\author[T.~Speltincx, E.~Laurikainen and H.~Salo]{T.~Speltincx$^{1}$\thanks{E-mail:Tom.Speltincx@oulu.fi}, E.~Laurikainen$^{1}$ and H.~Salo$^{1}$\\
$^{1}$Division of Astronomy, Dept. of Physical Sciences, University of Oulu, PO Box 3000, FIN-90014 University of Oulu, Finland}

\begin{document}

\date{Accepted 2007 October 3. Received 2007 October 1; in original form 2007 May 29}

\pagerange{\pageref{firstpage}--\pageref{lastpage}} \pubyear{2007}

\maketitle

\label{firstpage}

\begin{abstract}
\noindent

We use a gravitational bar torque method to compare bar strengths (the maximum tangential force normalized by radial force) in $B$ and $H$-band images of 152 galaxies from the Ohio State University Bright Spiral Galaxy Survey. Our main motivation is to check how much the difference in the rest-frame wavelength could affect comparisons of bar strengths in low and high redshift observations. Between these two bands we find an average bar strength ratio \QBH= $1.25$ which factor is nearly independent of the morphological type. We show that \QBH $> 1$ is mostly due to reduced bulge dilution of radial forces in the $B$-band. The bar torque method needs an estimate for the vertical scale height of the galaxy, based on the radial scale length of the disk and the galaxy's morphological type. Since these two might not always be possible to determine at high redshifts in a reliable manner, we also checked that similar results are obtained with vertical scale heights estimated from the radii corresponding to the $K$-band surface brightness of 20 mag/arcsec$^2$. Also, we made a simple test of the usability of the bar torque method at high redshifts by checking the effects of image degradation (nearest neighbour sampling without any adjustment of noise levels): we found that the estimated bar strengths varied by $\pm 10\%$ at most as long as the total extent of the bar was at least 10 pixels. Overall, we show that the gravitational bar torque method should provide a proficient tool for quantifying bar strengths also at high redshifts.

\end{abstract}


\section{Introduction}

Bars are important drivers of galaxy evolution, which highlights the importance of studying bar fractions and the properties of bars at a large range of redshifts. Bars generally have fairly old stellar populations, thus being well visible in the near-IR, where also the obscuration by dust is minimal. However, the near-IR images we obtain from the distant galaxies are redshifted so that at redshift $z \sim 1$ their rest-frame wavelength lies in the optical. Therefore, it is important to compare the optical and near-IR properties of bars in local galaxies.  For that purpose the $B$ and $H$-band images are very convenient, offering a wide separation in wavelengths. Namely, bars are still prominent in the $B$-band \citep{osu,whyte,sheth06}, where they can be identified in a fairly reliable manner, whereas, due to the Balmer break, bars typically disappear in the ultraviolet \citep{sheth}. On the other hand, the $H$-band is already fairly free of the effects of dust and thus offers a reliable tracer for the stellar mass distribution.

If bars were formed due to instabilities in the disk, one would expect a large fraction of barred galaxies at high redshifts in the framework of the hierarchical clustering models. This is because these models indicate dynamically colder disks in the distant Universe \citep*{NFW}. Also, interactions of disk galaxies were probably more frequent in the past \citep*[and references therein]{fer}, promoting the formation of bars \citep*[discussed in][]{elm}. Therefore, it was surprising to find a lack of bars in galaxies at redshift of $z>0.7$ (\citealp{abe}; \citealp{vdb96}; \citealp{vdb00}; \citealp*{vdb01}). Since this might be an artifact due to noise, poorer image resolution, and redshifted wavelength, \citet{vdb} degraded about 100 $B$-band images of the Ohio State University Bright Spiral Galaxy Survey ({\sc osubsgs}) to study these effects. By mimicking the characteristics of the images in the Hubble Deep Field,  using the Wide Field Planetary Camera 2 on HST, they created artificial images of those galaxies as they would appear in the $I$-band at a redshift $z=0.7$. They found that two-thirds of bars initially classified as strong bars (SB), were still detectable in the degraded images, whereas weak bars (SAB) would largely disappear. They concluded that while selection effects reduce the number of observed bars, they cannot completely explain the lack of bars at high redshift.

The early studies of bar frequencies were challenged by \citet{sheth}, who found similar bar frequencies at high and at low redshifts for very large bars. However, the number statistics were a severe concern, since only four bars were identified. These authors used galaxies in the Hubble Deep Field-North, observed at $V$, $I$ and $H$-bands, paying particular attention to the resolution limits due to which small bars cannot be detected at high redshifts. In two subsequent studies \citep{elm,jogee}, higher resolution images were used, based on the optical Advanced Camera for Surveys (ACS) at the Hubble Space Telescope. Both studies found a constant bar fraction in the redshift range $z = 0-1$. \citet{jogee} also found that it holds separately for the distinct intervals $z = 0-0.7$ and $z = 0.7-1.0$, using broad $V$ and $z$-bands to identify the bars. Taking into account the wavelengths used in their studies this constant bar fraction is not completely expected. Namely, if the true bar fraction is the same at all redshifts and bars are more prominent in the near-IR, one would expect a larger number of detected bars in the intermediate redshifts. There the rest-frame band shifting is not yet important, so that both strong and weak bars should be detected \citep{sheth06}. However, in the above studies the number statistics are still too poor (less than 300 galaxies spread over the entire redshift range) for any reliable discussion of how the bar fraction evolves as a function of redshift. Also, using ACS very deep images are required for detecting weak bars at high redshift.


The above studies have been superseded recently by \citet{sheth07}, using wide field {\sc cosmos} observations in $I$-band at $z < 0.84$, based on deeper images than used in the previous studies. Contrary to most previous studies they found that the fraction of bars declines rapidly with redshift. This drop of bars was also found to be more dramatic for strong bars. These studies of bar frequencies clearly show, besides the difficulty of measuring bar frequencies at high redshifts, also the importance of measuring the strengths of the bars at high redshifts in a reliable manner.

In this study we use the {\sc osubsgs} to compare bar strengths derived from the $B$ and $H$-band images of nearby galaxies, using a gravitational bar torque method. Besides bar strength, our method gives simultaneously also the length and the relative luminosity of the bar. The outline of this paper is as follows. In Sections \ref{sample} and \ref{method} we describe the details of our sample and the method used, and also address the factors which might affect the bar strength measurements at higher redshifts. We then show comparisons between $B$ and $H$-band bar strengths in Section \ref{highz}; the results are discussed in Section \ref{discussion}, and summarized in Section \ref{conclusions}.

\section{Sample}\label{sample}

We use the $B$ and $H$-band images from the {\sc osubsgs} \citep{osu}. The original {\sc osubsgs} sample consists of 205 spiral galaxies, selected from the Third Reference Catalogue of Bright Galaxies \citep{RC3}, with the following criteria: Hubble type index $0 \leq T \leq 9$, $B$ magnitude $m_B \leq 12$, galaxy diameter $D \leq 6\farcm 5$ and a declination in the range $-80 \degrees \leq \delta \leq +50 \degrees$.  In this study we use a subsample of the {\sc osubsgs}, with a restriction in inclination set to $i$ $<$ 65$\degrees$, limiting the number to 156 galaxies. The properties of this sample have been discussed in detail by \citet*{bls}. The requirement that both $B$ and $H$-band images must be available restricts our final number of galaxies to 152. Additionally, we leave out NGC 3338 in some of our calculations because of a bright saturated foreground star in the $B$-band. The pixel size in the $B$-band images varies in the range between 0.36 and 0.53 arcsec/pixel, whereas in the $H$-band images it varies between 1.11 and 1.50. In our sample of 152 galaxies, we find 102 galaxies with bars in the $H$-band, and 67 in the $B$-band. These bars are detected from studying the galaxy's Fourier density amplitude profiles. We describe the details in the next section.

\begin{table*}
\begin{minipage}{126mm}
\caption{Barred galaxies in our sample and their bar lengths and bar torque properties. The columns are the following: (1) NGC number; (2) Revised morphological type index from RC3; (3) $r_{K_{20}}$ radius from 2MASS catalogs; (4)-(5) bar lengths estimated by Fourier analysis in $B$- and $H$-band, respectively. The latter measurements are taken from \citet{lsbv}; (6) and (8) bar strength $Q_g$, measured using the $H$ and $B$-bands; (7) and (9) the radius of $Q_g$ in the same bands; (10) $Q_T$-value from the $B$-band, corresponding to $r_{Q_g}(H)$. \label{table}}
\begin{tabular}{cccccccccc}
\hline
Galaxy & T & $r_{K_{20}}$ & $barlen(B)$ & $barlen(H)$ & $Q_g(H)$ & $r_{Q_g}(H)$ & $Q_g(B)$ & $r_{Q_g}(B)$ & $Q_g(BH)$\\
(1) & (2) & (\arcsec) (3) & (\arcsec) (4) & (\arcsec) (5) & (6) & (\arcsec) (7) & (8) & (\arcsec) (9) & (10) \\
\hline
NGC 0150 & 3 & 56.3 & 25 & 27 & 0.47 & 24 & 0.55 & 25 & 0.55 \\
NGC 0210 & 3 & 54.1 & - & 46 & 0.08 & 34 & 0.10 & 33 & 0.10 \\
NGC 0289 & 4 & 65.8 & - & 20 & 0.21 & 13 & 0.45 & 3.7 & 0.30 \\
NGC 0428 & 9 & 42.5 & 44 & 45 & 0.24 & 44 & 0.39 & 44 & 0.39 \\
NGC 0578 & 5 & - & 18 & 20 & 0.21 & 8.1 & 0.26 & 7.8 & 0.26 \\
NGC 0613 & 4 & 118.2 & 98 & 104 & 0.40 & 68 & 0.53 & 72 & 0.52 \\
NGC 0685 & 5 & 66.7 & 21 & 21 & 0.46 & 3.5 & 0.49 & 9.4 & 0.43 \\
NGC 0864 & 5 & 61.8 & - & 26 & 0.36 & 20 & 0.48 & 21 & 0.48 \\
NGC 1073 & 5 & 57.7 & 37 & 38 & 0.62 & 23 & 0.75 & 19 & 0.68 \\
NGC 1087 & 5 & 59.5 & 10 & 18 & 0.59 & 4.5 & 0.76 & 2.4 & 0.68 \\
NGC 1187 & 5 & 86.1 & - & 29 & 0.21 & 36 & 0.33 & 37 & 0.33 \\
NGC 1241 & 3 & 56.5 & 20 & 30 & 0.26 & 20 & 0.30 & 24 & 0.28 \\
NGC 1300 & 4 & 131.8 & 85 & 87 & 0.55 & 66 & 0.82 & 66 & 0.82 \\
NGC 1302 & 0 & 80.0 & 21 & 26 & 0.08 & 22 & 0.08 & 22 & 0.08 \\
NGC 1317 & 1 & 63.1 & 49 & 58 & 0.09 & 38 & 0.10 & 35 & 0.10 \\
NGC 1350 & 2 & 100.1 & - & 81 & 0.26 & 64 & 0.29 & 75 & 0.28 \\
NGC 1385 & 6 & 56.1 & 10 & 9.3 & 0.57 & 3.5 & 0.64 & 5.3 & 0.62 \\
NGC 1493 & 6 & 66.6 & - & 23 & 0.36 & 10 & 0.63 & 11 & 0.63 \\
NGC 1559 & 6 & 89.7 & - & 17 & 0.33 & 5.8 & 0.52 & 10 & 0.42 \\
NGC 1617 & 1 & 115.7 & 16 & 22 & 0.29 & 3.3 & 0.35 & 1.2 & 0.32 \\
NGC 1637 & 5 & 76.6 & 22 & 23 & 0.20 & 17 & 0.18 & 15 & 0.17 \\
NGC 1703 & 3 & 51.4 & 10 & 11 & 0.86 & 7.8 & 0.10 & 10 & 0.09 \\
NGC 1808 & 1 & 131.1 & 86 & 87 & 0.29 & 75 & 0.59 & 67 & 0.55 \\
NGC 1832 & 4 & 51.6 & - & 17 & 0.20 & 12 & 0.21 & 11 & 0.20 \\
NGC 2139 & 6 & 43.8 & 16 & 17 & 0.41 & 3.3 & 0.46 & 5.8 & 0.38 \\
NGC 2207 & 4 & 72.5 & 35 & 46 & 0.33 & 29 & 0.75 & 29 & 0.75 \\
NGC 2442 & 4 & 126.4 & - & 93 & 0.67 & 78 & 0.99 & 76 & 0.98 \\
NGC 2559 & 4 & 94.7 & 27 & 33 & 0.33 & 23 & 0.43 & 28 & 0.41 \\
NGC 2566 & 2 & 90.2 & 62 & 72 & 0.33 & 52 & 0.40 & 53 & 0.40 \\
NGC 2964 & 4 & 58.1 & - & 30 & 0.32 & 20 & 0.32 & 20 & 0.32 \\
NGC 3059 & 4 & 116.6 & 20 & 20 & 0.61 & 3.3 & 0.59 & 5.8 & 0.58 \\
NGC 3166 & 0 & 81.7 & - & 45 & 0.17 & 29 & 0.15 & 25 & 0.15 \\
NGC 3227 & 1 & 92.6 & 82 & 75 & 0.16 & 56 & 0.34 & 69 & 0.30 \\
NGC 3261 & 3 & 61.6 & 23 & 28 & 0.21 & 19 & 0.27 & 21 & 0.26 \\
NGC 3275 & 2 & 68.7 & - & 41 & 0.19 & 23 & 0.22 & 28 & 0.19 \\
NGC 3319 & 6 & 45.5 & 36 & 38 & 0.55 & 14 & 0.83 & 14 & 0.83 \\
NGC 3338 & 5 & 91.9 & - & 23 & 0.22 & 4.5 & 0.29 & 1.5 & 0.20 \\
NGC 3504 & 2 & 65.3 & 59 & 60 & 0.30 & 29 & 0.45 & 31 & 0.43 \\
NGC 3507 & 3 & 88.6 & 21 & 23 & 0.18 & 20 & 0.20 & 19 & 0.19 \\
NGC 3513 & 5 & 71.7 & 27 & 28 & 0.54 & 14 & 0.81 & 16 & 0.80 \\
NGC 3583 & 3 & 47.6 & 22 & 23 & 0.25 & 17 & 0.31 & 23 & 0.28 \\
NGC 3593 & 0 & 101.0 & - & 15 & 0.10 & 17 & 0.10 & 23 & 0.05 \\
NGC 3675 & 3 & 120.7 & - & 30 & 0.15 & 4.5 & 0.20 & 1.4 & 0.14 \\
NGC 3681 & 4 & 38.8 & 12 & 15 & 0.22 & 4.5 & 0.20 & 3.4 & 0.20 \\
NGC 3686 & 4 & 87.2 & 15 & 18 & 0.28 & 11 & 0.22 & 14 & 0.20 \\
NGC 3726 & 5 & 103.2 & - & 30 & 0.21 & 26 & 0.22 & 28 & 0.22 \\
NGC 3887 & 4 & 78.5 & - & 41 & 0.21 & 32 & 0.24 & 36 & 0.24 \\
NGC 4027 & 8 & 65.5 & 16 & 20 & 0.62 & 3.3 & 0.64 & 1.9 & 0.63 \\
NGC 4051 & 4 & 102.6 & 41 & 45 & 0.29 & 56 & 0.33 & 48 & 0.34 \\
NGC 4123 & 5 & 65.9 & - & 53 & 0.43 & 38 & 0.66 & 37 & 0.66 \\
NGC 4136 & 5 & 53.8 & - & 15 & 0.13 & 11 & 0.13 & 10 & 0.13 \\
NGC 4145 & 7 & 62.8 & 15 & 20 & 0.36 & 4.5 & 0.28 & 1.5 & 0.29 \\
NGC 4151 & 2 & 88.2 & 97 & 98 & 0.13 & 65 & 0.20 & 66 & 0.20 \\
NGC 4293 & 0 & 117.1 & 71 & 68 & 0.36 & 50 & 0.40 & 42 & 0.39 \\
NGC 4303 & 4 & 105.8 & - & 30 & 0.27 & 41 & 0.45 & 39 & 0.45 \\
\hline
\end{tabular}
\end{minipage}
\end{table*}

\addtocounter{table}{-1}

\begin{table*}
\begin{minipage}{126mm}
\caption{Continued}
\begin{tabular}{cccccccccc}
\hline
Galaxy & T & $r_{K_{20}}$ & $barlen(B)$ & $barlen(H)$ & $Q_g(H)$ & $r_{Q_g}(H)$ & $Q_g(B)$ & $r_{Q_g}(B)$ & $Q_g(BH)$\\
& & (\arcsec) & (\arcsec) & (\arcsec) & & (\arcsec) & & (\arcsec) & \\
\hline
NGC 4314 & 1 & 110.1 & 71 & 75 & 0.46 & 53 & 0.45 & 53 & 0.45 \\
NGC 4394 & 3 & 97.4 & 39 & 45 & 0.28 & 32 & 0.27 & 33 & 0.27 \\
NGC 4450 & 2 & 112.0 & - & 30 & 0.12 & 41 & 0.11 & 40 & 0.11 \\
NGC 4457 & 0 & 62.1 & 43 & 45 & 0.09 & 32 & 0.11 & 29 & 0.10 \\
NGC 4487 & 6 & 67.3 & 18 & 22 & 0.17 & 7.8 & 0.81 & 1.2 & 0.24 \\
NGC 4490 & 7 & 117.4 & - & 18 & 0.33 & 7.5 & 0.54 & 2.6 & 0.45 \\
NGC 4496 & 9 & - & 22 & 26 & 0.36 & 7.5 & 0.55 & 11 & 0.51 \\
NGC 4527 & 4 & 141.0 & - & 83 & 0.18 & 77 & 0.25 & 74 & 0.24 \\
NGC 4548 & 3 & 105.0 & 61 & 68 & 0.36 & 53 & 0.36 & 52 & 0.36 \\
NGC 4579 & 3 & 120.0 & 46 & 45 & 0.20 & 32 & 0.22 & 33 & 0.22 \\
NGC 4593 & 3 & 80.6 & - & 61 & 0.32 & 43 & 0.31 & 42 & 0.31 \\
NGC 4618 & 9 & 73.1 & 26 & 33 & 0.36 & 11 & 0.64 & 12 & 0.64 \\
NGC 4643 & 0 & 92.5 & 62 & 68 & 0.27 & 41 & 0.27 & 41 & 0.27 \\
NGC 4647 & 5 & 86.0 & 11 & 12 & 0.24 & 11 & 0.37 & 10 & 0.37 \\
NGC 4651 & 5 & 69.1 & 14 & 23 & 0.17 & 4.5 & 0.22 & 1.8 & 0.16 \\
NGC 4654 & 6 & 96.4 & 15 & 20 & 0.14 & 11 & 0.31 & 10 & 0.31 \\
NGC 4665 & 0 & 110.1 & 53 & 60 & 0.27 & 38 & 0.25 & 34 & 0.25 \\
NGC 4691 & 0 & 71.5 & 45 & 45 & 0.59 & 11 & 0.79 & 13 & 0.73 \\
NGC 4699 & 3 & 100.9 & - & 14 & 0.08 & 10 & 0.09 & 12 & 0.09 \\
NGC 4781 & 7 & 69.7 & - & 39 & 0.44 & 14 & 0.60 & 15 & 0.59 \\
NGC 4900 & 5 & 64.5 & 14 & 18 & 0.34 & 14 & 0.59 & 18 & 0.49 \\
NGC 4902 & 3 & 49.1 & - & 22 & 0.28 & 17 & 0.34 & 17 & 0.34 \\
NGC 4930 & 3 & 58.3 & 43 & 44 & 0.21 & 32 & 0.27 & 35 & 0.26 \\
NGC 4939 & 4 & 62.7 & - & 17 & 0.20 & 3.5 & 0.22 & 1.2 & 0.17 \\
NGC 4995 & 3 & 57.1 & 18 & 23 & 0.28 & 20 & 0.27 & 25 & 0.25 \\
NGC 5005 & 4 & 130.8 & - & 45 & 0.15 & 26 & 0.10 & 37 & 0.06 \\
NGC 5101 & 0 & 97.8 & 66 & 70 & 0.22 & 43 & 0.25 & 45 & 0.25 \\
NGC 5334 & 5 & 51.1 & 16 & 18 & 0.36 & 11 & 0.39 & 12 & 0.38 \\
NGC 5483 & 5 & 68.5 & - & 13 & 0.16 & 7.8 & 0.25 & 8.9 & 0.25 \\
NGC 5643 & 5 & 99.5 & - & 46 & 0.42 & 34 & 0.48 & 35 & 0.48 \\
NGC 5701 & 0 & 74.3 & 47 & 50 & 0.16 & 28 & 0.17 & 30 & 0.17 \\
NGC 5713 & 4 & 54.0 & - & 30 & 0.43 & 4.5 & 0.77 & 6.0 & 0.66 \\
NGC 5850 & 3 & 79.2 & 74 & 90 & 0.33 & 59 & 0.36 & 62 & 0.36 \\
NGC 5921 & 4 & 75.4 & 56 & 53 & 0.42 & 47 & 0.50 & 52 & 0.49 \\
NGC 5962 & 5 & 49.6 & 14 & 15 & 0.13 & 14 & 0.12 & 12 & 0.10 \\
NGC 6221 & 5 & - & 35 & 41 & 0.44 & 29 & 0.52 & 28 & 0.52 \\
NGC 6300 & 3 & 116.8 & 53 & 46 & 0.19 & 34 & 0.29 & 32 & 0.29 \\
NGC 6384 & 4 & 99.6 & - & 33 & 0.11 & 17 & 0.10 & 18 & 0.10 \\
NGC 6782 & 1 & 45.4 & 45 & 46 & 0.17 & 24 & 0.22 & 25 & 0.22 \\
NGC 6902 & 3 & 53.0 & - & 17 & 0.08 & 3.5 & 0.09 & 1.2 & 0.06 \\
NGC 7418 & 6 & 66.2 & - & 15 & 0.20 & 15 & 0.29 & 18 & 0.26 \\
NGC 7479 & 5 & 87.9 & 59 & 60 & 0.70 & 41 & 0.88 & 37 & 0.83 \\
NGC 7552 & 2 & 76.4 & 74 & 70 & 0.42 & 45 & 0.64 & 45 & 0.64 \\
NGC 7582 & 2 & 115.2 & 82 & 93 & 0.44 & 55 & 0.52 & 52 & 0.52 \\
NGC 7723 & 3 & 68.4 & 20 & 24 & 0.35 & 17 & 0.28 & 17 & 0.28 \\
NGC 7727 & 1 & 63.2 & 15 & 27 & 0.16 & 4.5 & 0.16 & 3.4 & 0.15 \\
NGC 7741 & 6 & 55.2 & 44 & 53 & 0.72 & 11 & 0.96 & 15 & 0.88 \\
\hline
\end{tabular}
\end{minipage}
\end{table*}

\section{Method and measurements}\label{method}

We use a gravitational bar torque method described by \citet{ls02}. Bar strengths are estimated from the magnitude of the non-axisymmetric gravitational perturbation compared to the mean axisymmetric radial force field. The relative strength of the perturbation at each radius $r$ is defined as the ratio of the maximum tangential force $F_T^{max}(r)$ to the azimuthally averaged mean radial force $F_R(r)$:

\begin{equation}
Q_T(r)=\frac{F_T^{max}(r)}{\langle F_R(r)\rangle}.
\end{equation}

\noindent To get a single measure of bar strength, we use the maximum of the $Q_T(r)$-profile in the bar region, denoted by $Q_g$. The radius where this maximum occurs, is denoted by $r_{Q_g}$. The method uses an exponential model for the vertical density profile and an empirical relation between the morphological type and the radial to vertical scale parameter ratio, $h_R/z_0$, \citep{degrijs} to estimate the vertical thickness. The refined method \citep{lsbv} also allows a bulge correction, in which case the different 3D distribution of bulge and disk components is taken into account in the calculation of the forces.

To study bar strengths with as little interference from the spiral arms as possible, \citet*{bbk} developed a technique for separating the bar and spiral contributions to the Fourier density amplitude profiles. In this study no such bar-spiral separation is made, mainly because it is more difficult to apply the method at high redshift. Also, the effect of this correction turns out to be small with respect to the mean $Q_g$-values in the Hubble type-index bins \citep{laur07}.

We measure $Q_g$ both using the $B$ and $H$-band images, and the results are collected in Table \ref{table}. Note that the table contains only the barred galaxies from our sample. The classification is based on Fourier analysis made in the $H$-band, which classifications are given in \citet*{lsb}. Here no bulge correction was applied, mainly because doing such a correction at high redshift is not always straightforward. Therefore, the effect of possible artificial bulge stretching while deprojecting the images was controlled by excluding possible central peaks of $Q_T$-values in the bulge dominated region. In the $H$-band we compared our $Q_g$-values without applying the bulge correction with the values obtained by \citet{lsbv} where that correction was applied. We found that although a bulge correction might be important for some individual galaxies, it does not affect the mean $Q_g$-values in the Hubble type-index bins \citep{lsbv}.

A similar test was made also in the $B$-band, by measuring the bulge-corrected $Q_g$-values for 26 randomly chosen galaxies. The bulge correction was made by applying the 2D multi-component decomposition code (\citealp{lsbv}; \citealp*{lsb2}). The method uses a S\'ersic function for the bulge, an exponential disk, and a Ferrers or S\'ersic function for the bar/oval. In order to obtain robust bulge and disk parameters, a two-component fit was made first. Using this fit for the initial parameters we then fitted the bar, which is a tricky task in the $B$-band, as part of the bar can be obscured by dust and by recent star formation. For this reason, we had to occasionally fix the bar radius in the fit, based on visual inspection of the image. Having found an acceptable decomposition, the derived bulge model is then subtracted from the original image, the image is deprojected to a face-on orientation, after which the bulge is added back and the forces are calculated.  We compared the $Q_g$-values before and after bulge correction, and found an average difference of 1\%. It is therefore safe to say that not making the bulge correction will barely influence our results.

The length of a bar can be estimated by looking at the phases of the $m=2$ and $m=4$ components of the Fourier density amplitudes: the phase is assumed to be maintained nearly constant in the bar region. For the $H$-band we used the bar lengths derived by \citet{lsbv}. For the $B$-band we derived them in this study, using a similar method, and the resulting values (denoted by $barlen$) can be found in Table \ref{table}. However, in the $B$-band the length of the bar could not always be estimated by the Fourier method, most probably due to the effects of dust that make the interpretation of the phase profile more complicated. Nevertheless, even for these cases one can still obtain the $B$-band $Q_T(r)$ profile and its maximum value in the bar region, $Q_g(B)$.

In Fig. \ref{fig:sequence} we show examples of galaxy images and the $Q_T$-profiles for different morphological types (figures for the rest of the galaxies are available in the electronic version). The left column shows the $B$-band images, whereas the right column shows the $H$-band images in the same scale and orientation. The center column shows the $Q_T$-profiles in both bands (full and dashed line for $B$ and $H$, respectively) and the bar lengths are indicated by a vertical line. It is clear from this figure that the $B$-band profile is generally less smooth due to the influence of the spiral arms. For the early-type galaxies, particularly for those with strong bars, the bar torques derived from the $B$ and $H$-band images are generally very similar, whereas for later type galaxies the $B$-band images seem to give slightly stronger bar strengths.

\begin{figure*}
\includegraphics[height=0.9\textheight]{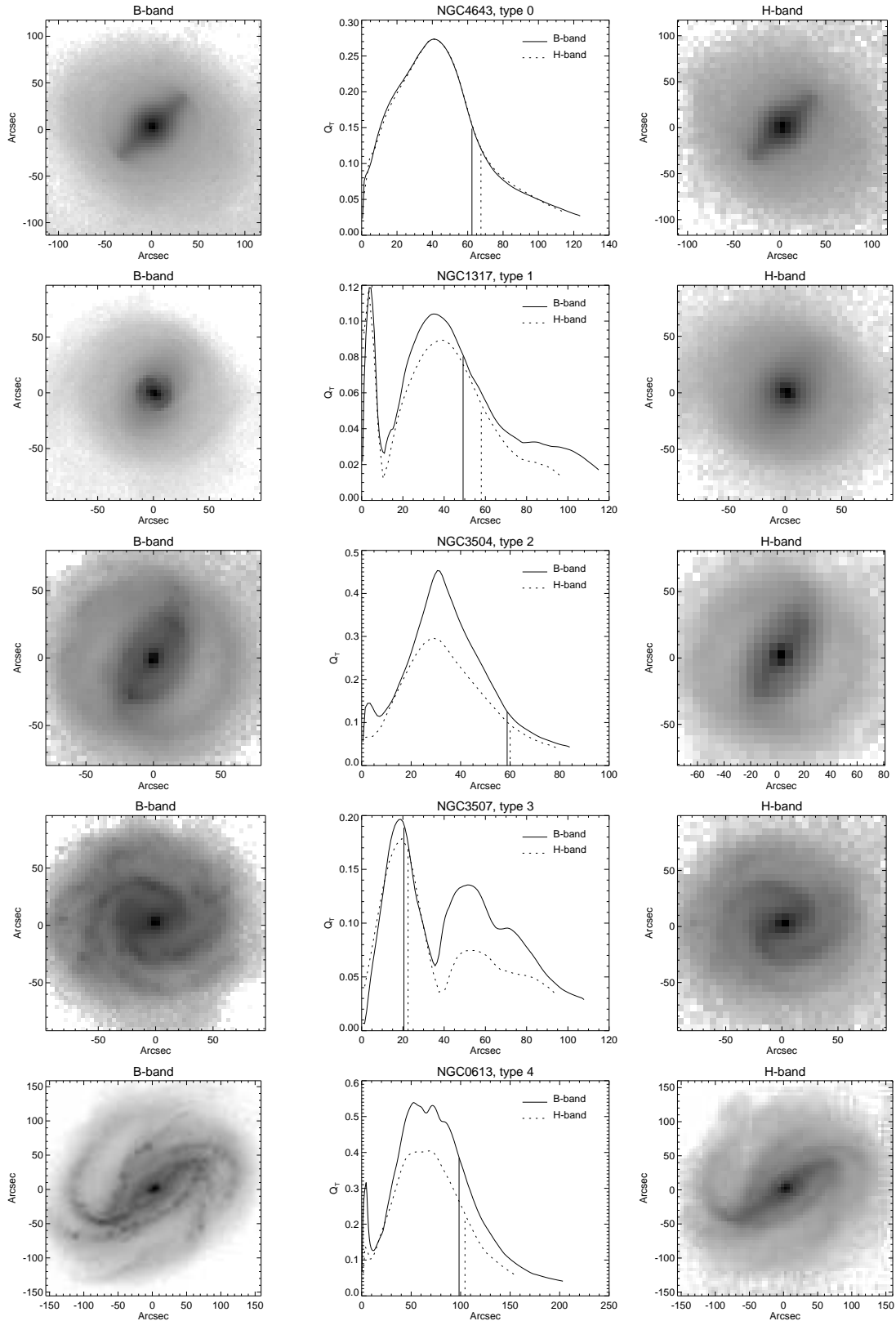}
\caption{A journey through the morphological types. The left column shows the $B$-band images, while the right column shows the $H$-band images. The center column shows the $Q_T$-profiles for the galaxies for $B$ and $H$-band in full and dashed lines respectively. The bar lengths in both bands are marked with a vertical line.\label{fig:sequence}}
\end{figure*}

\addtocounter{figure}{-1}

\begin{figure*}
\includegraphics[height=0.9\textheight]{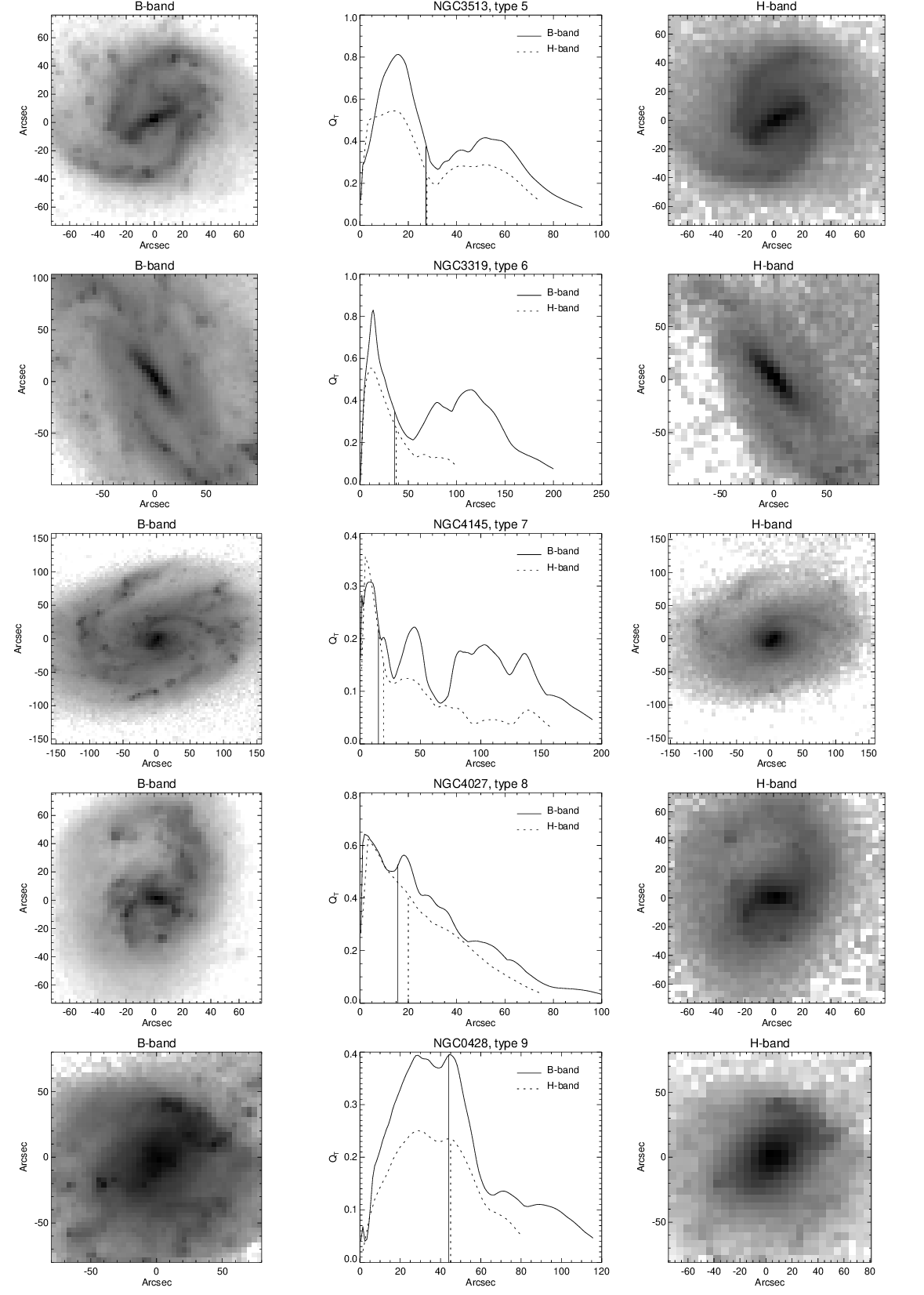}
\caption{(cont.)}
\end{figure*}

\section{The bar torque method at high redshift}\label{highz}
\subsection{Comparing bar strengths}\label{comparisons}

\begin{figure*}
\includegraphics[height=0.6\textheight]{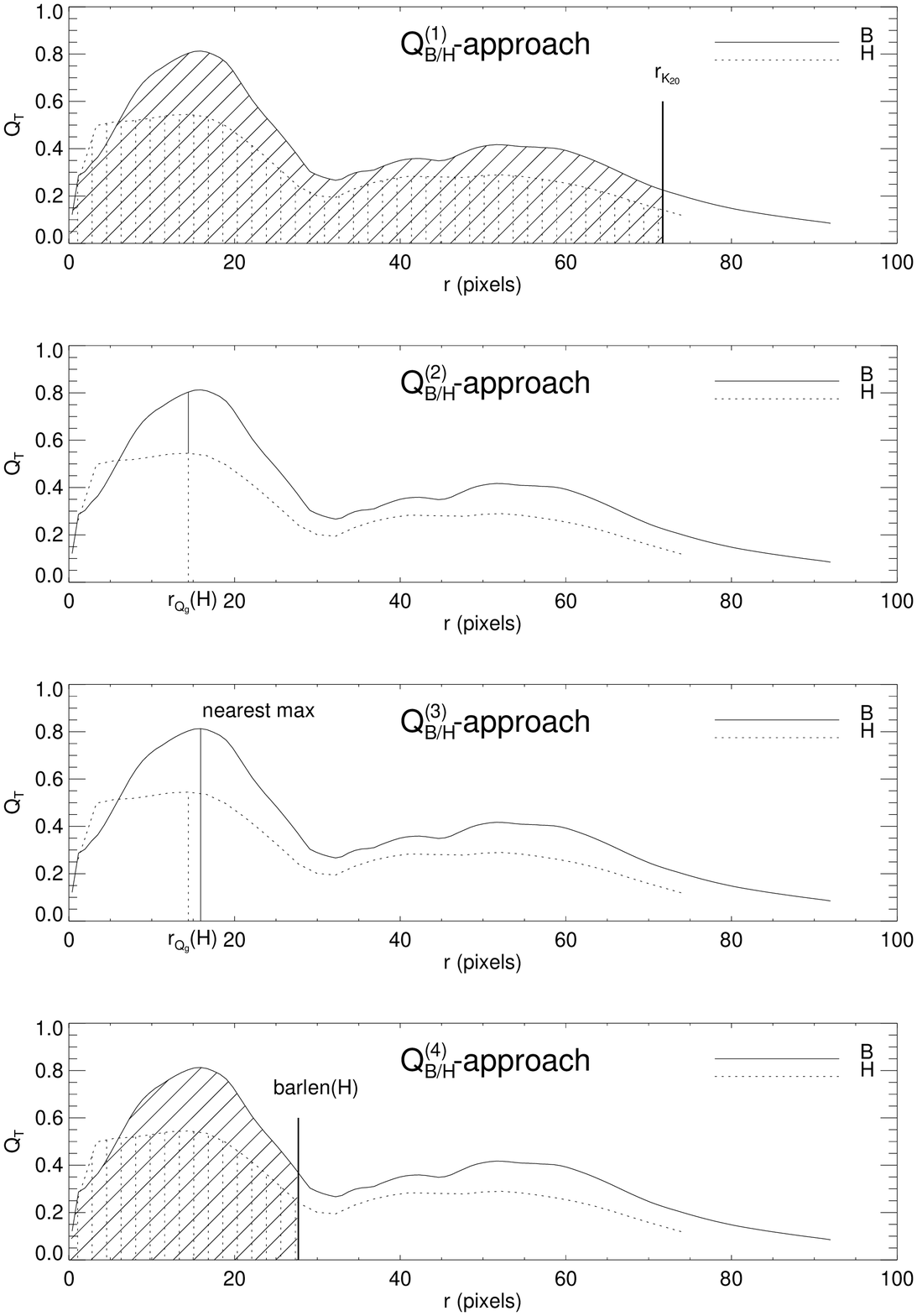}
\caption{Visualization of the four methods for NGC 3513. From top to bottom: \Z, where we use the area below the $Q_T$ curves up to $r_{K_{20}}$; \ZZ, where we use the maximum value of $Q_T$ in the bar region, i.e. $Q_g$, of the $H$-band image and the $Q_T$ value of the $B$-band image at the same radius; \ZZZ, similar to the second method, but we now take the maximum $Q_T$ value in the $B$-band image in the bar region as well; \ZZZZ, where we use the area below the $Q_T$ curves, taking the bar length in the $H$-band image to be the maximum radius.\label{fig:Qs}}
\end{figure*}

In order to compare bar strengths in $B$ and $H$-bands we use four approaches, all based on the gravitational bar torque method. They are graphically illustrated in Fig. \ref{fig:Qs} for one typical case (NGC 3513).  The main idea is to compare bar strengths derived from the $B$ and $H$-band images, either using a single number for the bar strength (maximum of $Q_T(r)$ in the bar region), or by integrating over the area under the $Q_T(r)$-profile. The ratio of the obtained bar strengths in the two wavelengths is then denoted by \Zi, where $i=1,2,3,4$ for the different cases. 

In the first case the averages are calculated using the limiting integration radius of $r_{K_{20}}$\footnote{We obtained the values for $r_{K_{20}}$ from the 2 Micron All Sky Survey (2MASS)}, corresponding to the radial distance where the $K_s$-band surface brightness is 20 $mag/arcsec^2$. The ratio $\langle Q_T(B)\rangle/\langle Q_T(H) \rangle$ will be denoted by \Z. 
In the second approach we use a single number, $Q_g$, to measure the bar strengths in the $H$-band. In the $B$-band we take the $Q_T$-value at the same radius (the ratio of $Q_T$'s is denoted by \ZZ). The third approach is quite similar to the second one, but differs in the fact that in the $B$-band we now take the real $Q_T$ maximum closest to the radius where the maximum $Q_T$ in the $H$-band appears (the ratio is denoted by \ZZZ).
Our fourth approach uses again the area below the $Q_T$-profiles. This time, though, we take the integration cut-off at the radius corresponding to the length of the bar as estimated from the $H$-band image (the ratio is denoted by \ZZZZ). Compared to \Z, the advantage is the elimination of the possible influence of spiral arms.

\begin{figure*}
\includegraphics[width=0.7\textwidth]{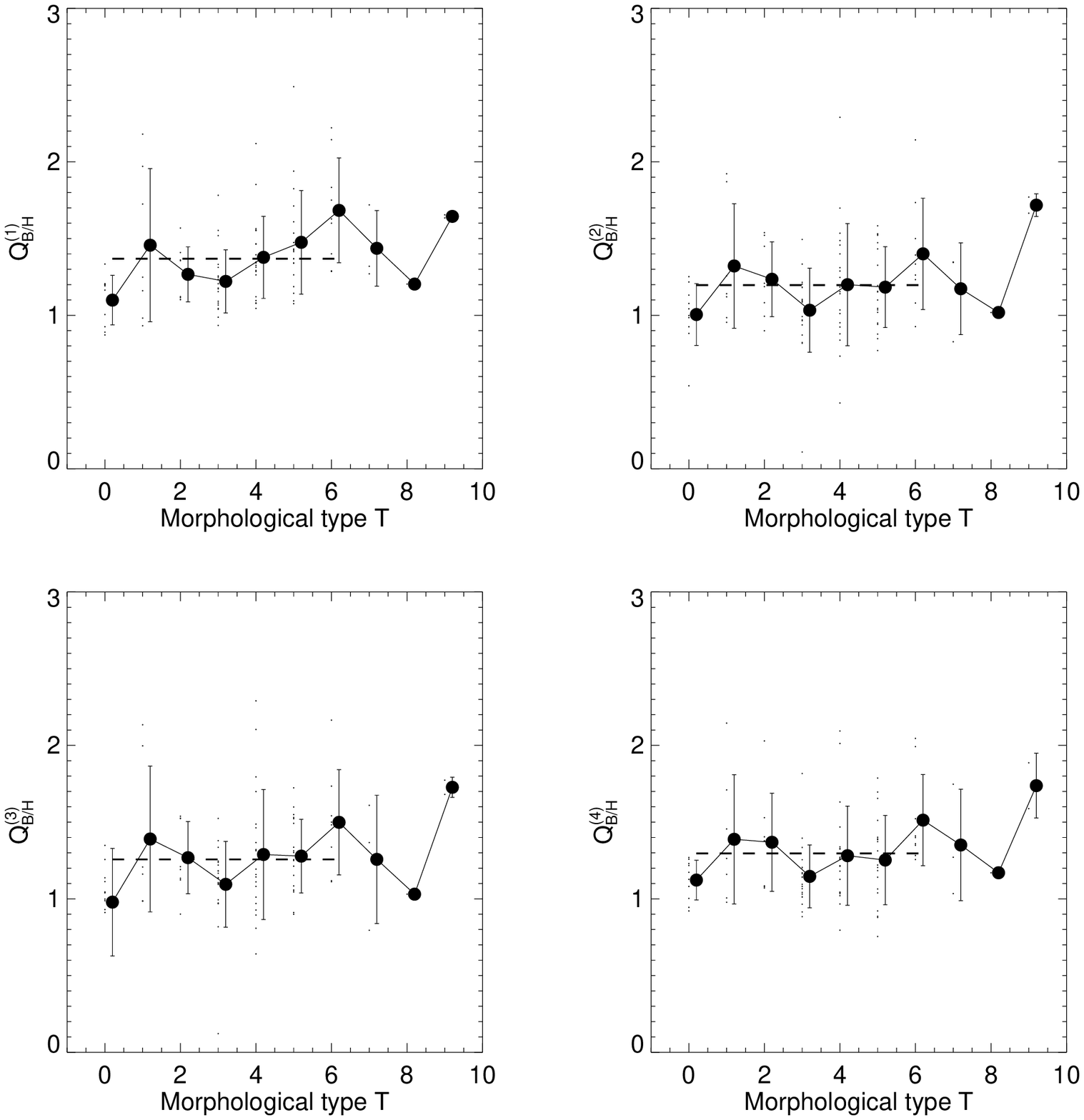}
\caption{\Zi versus morphological type. For \Z we see a slight increase as we go to later types, which is an effect of the prominence of spiral arms in those types. For the other three methods, the \Zi's are fairly independent of the type except for T $> 6$, where the number of galaxies and hence the reliability of the statistics drop considerably. The dashed line marks the average value through those seven types. The length of the vertical bars indicates the standard deviation for each type bin.
\label{fig:QvsT}}
\end{figure*}

In Figure \ref{fig:QvsT} we show how the \Zi's vary with the morphological type. These plots are based on 99 barred galaxies, omitting the three galaxies for which no $K_s$-band luminosities were found from 2MASS. For \Z there is a slight increase as we move to later types (\Z$=1.1-1.7$ over types 0 through 6, with an average of 1.37), most probably related to the prominence of spiral arms well outside the bar region in those types. For the other three approaches, we get a more constant behavior in the Hubble sequence. Since from type 7 onward we have only 1 or 2 galaxies per type-index bin, statistics there are not reliable.

Using all galaxies in which a bar was identified in the $H$-band, but limiting to the morphological types T $= 0 - 6$, we find an average bar strength ratio \Zi = 1.20, 1.26 and 1.30 for $i=2,3$ and $4$, respectively. These values are indicated with a dashed line in Fig. \ref{fig:QvsT}. If we take into account only those 64 galaxies for which bars were found in both wavelengths, \ZZ changes from 1.20 to 1.22, while \ZZZ and \ZZZZ (and also \Z) remain the same. In conclusion, it is safe to say that the \QBH ~ ratio is fairly independent of the morphological type, so that we can adopt an average value of \QBH $= 1.25$ for converting bar strengths between $B$ and $H$-bands. Therefore, while measuring band-shifted optical bar strengths in galaxies at $z \ge 1$ using near-IR images, corresponding true near-IR bar strengths can be estimated for all morphological types by applying this correction. This makes it possible to directly compare bar strengths at low and at high redshifts.

Considering that bars are typically easier to detect in the $H$-band, it may be surprising to find larger bar strengths in the $B$-band. In principle, this might indicate that the density contrast of bars is bigger in the $B$-band compared to that in the $H$-band, leading to larger apparent tangential forces. However, comparison of the Fourier density amplitudes indicates that the density variations associated with the bars are about the same ($\langle A_2^B/A_2^H\rangle \approx 1.02$).

Another possibility is that the bulge dilution effect is smaller in the $B$-band compared to the $H$-band, so that the apparent radial forces are reduced. To investigate this further, we make a crude analytical estimate of the bar strength ratio. We approximate the tangential force as $F_T \propto A_2 d$, where $A_2$ is the Fourier density amplitude of the $m=2$ term and $d$ is the disk mass. The radial force is proportional to the bulge and the disk mass, but both don't necessarily contribute a similar amount:

\begin{equation}
\frac{F_b}{F_d}=f\frac{b}{d}
\end{equation}

\noindent where $F$ is a radial force, with subscripts $b$ and $d$ for bulge and disk, respectively. The mass of the bulge component is denoted by $b$. Here, $f$ is a factor of proportionality that depends on the distance from the galactic center (i.e. the distance where the $Q_T$ maximum occurs). This way, we can approximate the bar strength ratio, calculated in $B$ and $H$-bands by:

\begin{eqnarray}
Q_{B/H} & \equiv & \frac{Q_T^B}{Q_T^H} \\ & \approx & \frac{d^B
A_2^B}{d^B + f b^B}\frac{d^H + f b^H}{d^H A_2^H} \\ & = &
\frac{A_2^B}{A_2^H}\frac{1+f~(\frac{b}{d})^H}{1+f~(\frac{b}{d})^B}
\label{dil}
\end{eqnarray}

\begin{figure*}
\includegraphics[width=0.7\textwidth]{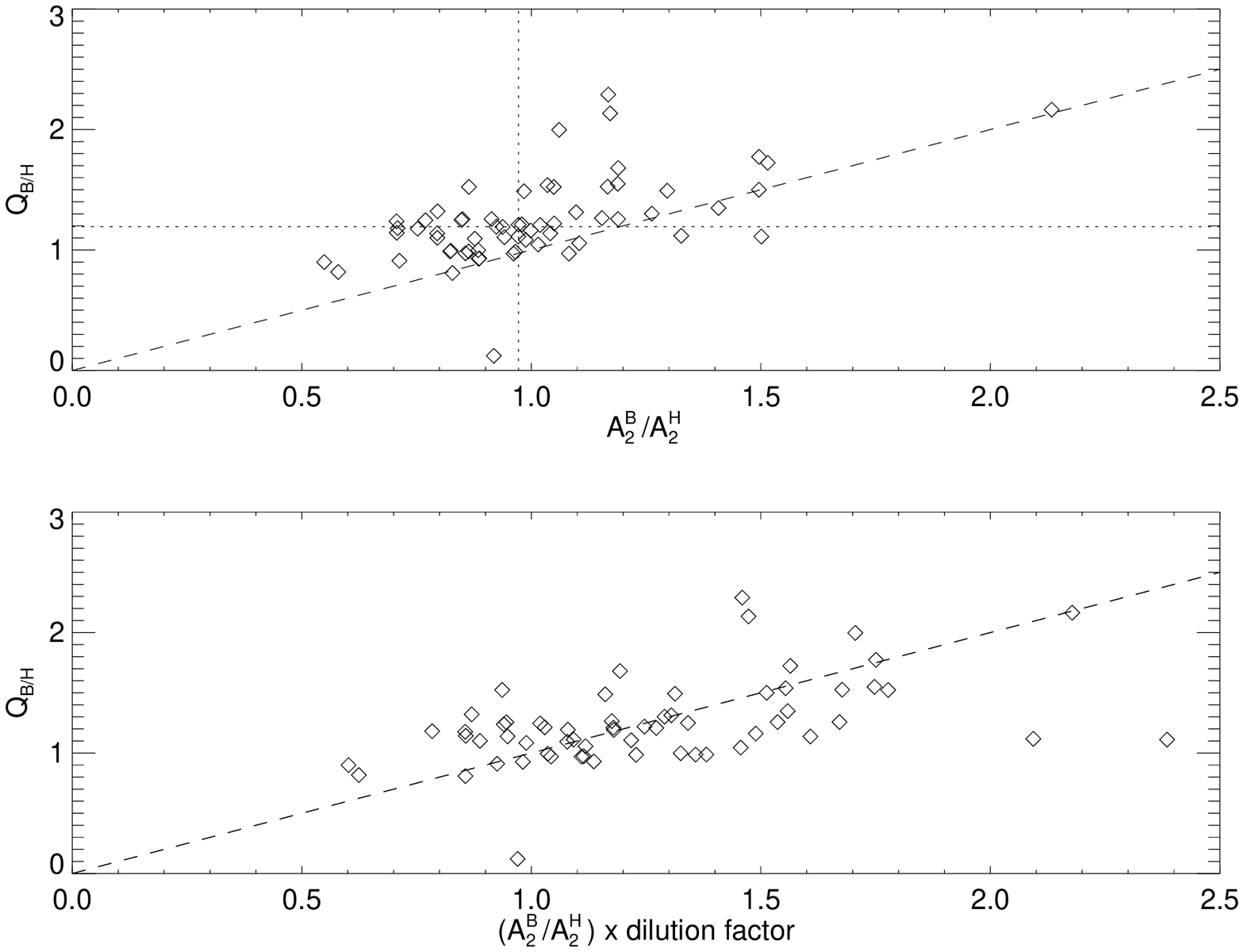}
\caption{The influence of bulge dilution. The top panel shows the relation between the bar strength ratio and the density contrast. The dashed lines indicate the median values. The bottom panel takes into account the different bulge-to-disk ratios in the two bands. For a full description, see Section \ref{comparisons}.\label{dilution}}
\end{figure*}

\noindent In case the bulge-to-disk mass ratio $b/d$ would be the same in both bands, the second factor on the right in Eq. (\ref{dil}) becomes unity, and the bar strength ratio simply equals the ratio of Fourier density amplitudes. We show the relation between $Q_{B/H}$ and $A_2^B/A_2^H$ in the upper panel of Fig. \ref{dilution}. It is clear from this plot that we cannot simply ignore the bulges. Since we do not have $(b/d)^B$ for all our sample galaxies, we make a further approximation like:

\begin{equation}
\frac{Q_T^B}{Q_T^H} \approx \frac{A_2^B}{A_2^H}
\frac{1+f(\frac{b}{d})^H}{1+f\left\langle\frac{(b/d)^B}{(b/d)^H}\right\rangle(\frac{b}{d})^H}\label{dil2}
\end{equation}

\noindent where the brackets denote a median, taken from the 26 galaxies for which we have decompositions in both bands. For this median, we find a value of $\sim 0.41$, consistent with bulges being red with respect to disks. To calculate an appropriate value for $f$, we approximate the bulge by a point-mass, while the disk is represented by an exponential with scale length $h_R$. Note that $r_{Q_g}$ occurs on average at about one exponential scale length: for $r/h_R \approx 1$ we have $f \sim 3$. The relation from Eq. (\ref{dil2}) with this value for $f$ is shown in the bottom panel of Fig. \ref{dilution}. There is a clear improvement, confirming the fact that \QBH $> 1$ is mostly due to reduced bulge dilution in the $B$-band.

\begin{figure*}
\includegraphics[width=0.7\textwidth]{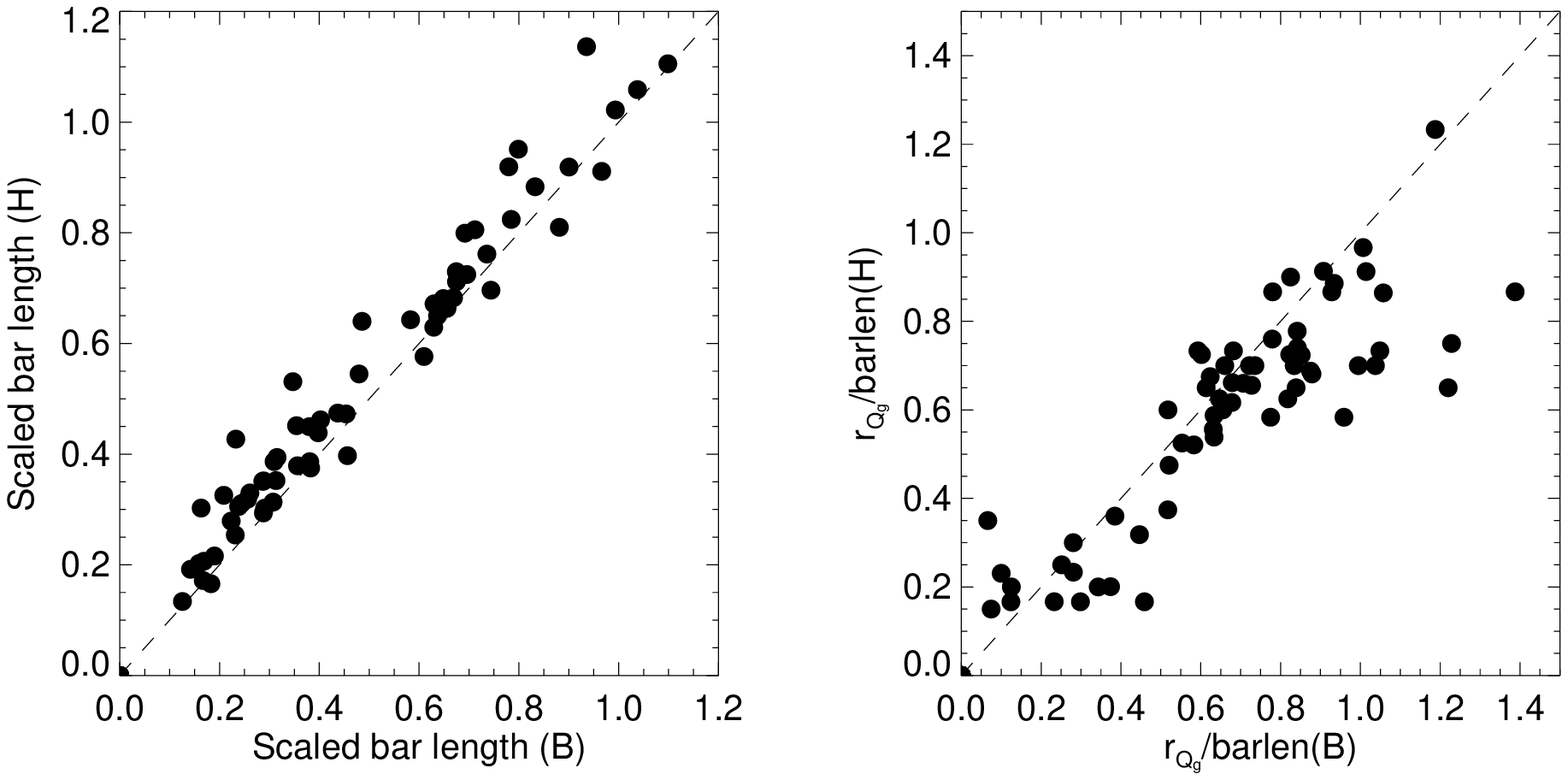}
\caption{The left panel of this figure shows a tight correlation between the bar lengths, scaled by $r_{K_{20}}$, in both bands. The right panel shows the correlation between $r_{Q_g}$ compared to the bar length, in both bands. The dashed lines indicate the unit slope.\label{fig:bars}}
\end{figure*}

The bar lengths in the two wavelengths are very similar, the one in the $B$-band usually being the slightly shorter one. The left panel of Fig. \ref{fig:bars} shows a tight correlation between the bar lengths in the two bands, the lengths being normalized to $r_{K_{20}}$. The radius of the maximum bar {$Q_T$}, $r_{Q_g}$, is well correlated with the length of the bar (estimated from the Fourier phases), 
in agreement with \citet*{lsr02}. In the right panel of Fig. \ref{fig:bars} we show $r_{Q_g}$ relative to the bar length, compared in the two bands. The outlying points are due to underestimated bar lengths in the $B$-band, most likely because part of the bar is obscured by dust. On the other hand, the maximum of the $Q_T$-profile still appears at the 'right' place, causing the ratio of these two to be overestimated.

\subsection{Estimating $z_0$ through $r_{K_{20}}$}\label{ztest}

\begin{figure*}
\includegraphics[width=0.8\textwidth]{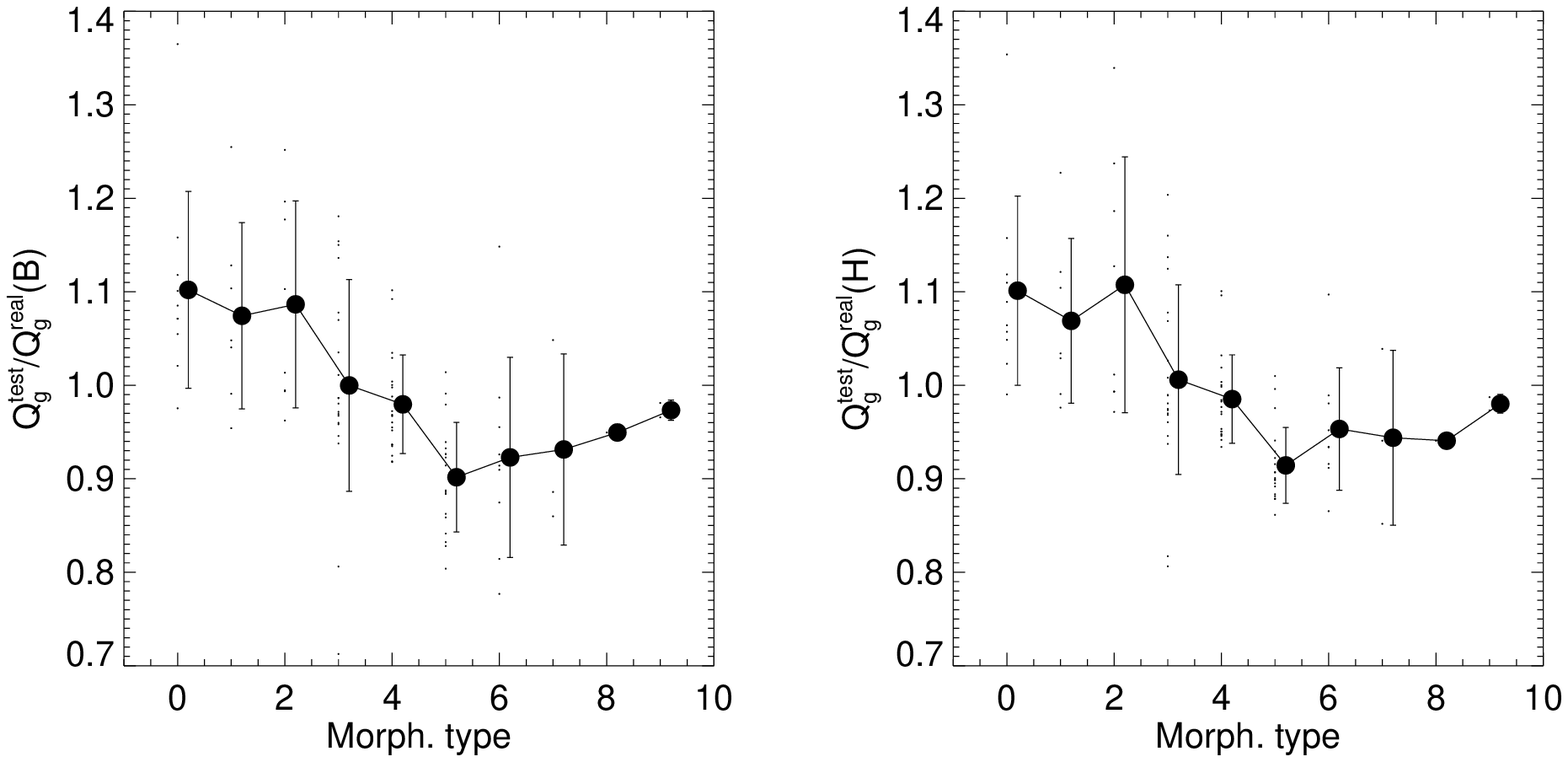}
\caption{Change of $Q_g$ values when $z_0$ is estimated from $r_{K_{20}}$ rather than $h_R$. A clear type dependence is noticeable. For early types the bar strengths will be overestimated, whereas for late types they will be underestimated.
\label{fig:ztest}}
\end{figure*}

The gravitational bar torque method makes explicit use of the radial scale length of the disk, $h_R$, to obtain the vertical scale height, $z_0$, through the empirical relation by \citet{degrijs}, where $z_0/h_R$ depends of the morphological type. We investigate here whether we can also get similar results without using this relation, by estimating the vertical scale height by means of the $r_{K_{20}}$ radius. We calculate the bar strengths again for the galaxies in our sample using the following approximation for $z_0$:

\begin{equation}
z_0 = \frac{z_0}{r_{K_{20}}}\;r_{K_{20}} \approx
\left\langle\frac{z_0}{r_{K_{20}}}\right\rangle r_{K_{20}} \label{z0}
\end{equation}

\noindent Here the ratio $\left\langle z_0/r_{K_{20}}\right\rangle$ is calculated using the original $z_0$ values estimated via the \citet{degrijs} relation. With this approximation, we make thin disks (mainly in late-type galaxies) a bit thicker, and thick disks (mainly in early-type galaxies) a bit thinner. As an average value for all morphological types in our sample we find $\left\langle z_0/r_{K_{20}}\right\rangle = 0.08$. This is by no means a universal value, but could be used for spirals. If one was to study S0 galaxies, another value would most likely be more suitable. In Fig. \ref{fig:ztest} we show how much the bar strength is affected when using this approximation. In both bands we compare the ratio of bar strengths derived using the above approximation, and the original method where the radial scale length was used for estimating the vertical thickness. There is a clear type dependence visible, which is no surprise since $z_0/h_R$ depends on the morphological type. Thus, estimating $z_0$ through $r_{K_{20}}$ implies that bar strengths for the early type galaxies would be overestimated, whereas for late type galaxies they would be underestimated, but by no more than 10-15$\%$. So, within these errors, the $Q_g$-method is still useful even if the morphological type and/or radial scale length can not be determined. Also note that the effect of this approximation is similar in both bands.

\subsection{The bar torque method using degraded images}\label{degrade}

So far we have tested our method on images from a local galaxy sample. In order to test whether the method is valid also for images with lower resolution, like those obtained at higher redshifts, we degraded the images of our sample. As this is not supposed to be a test for any particular telescope or survey, we don't follow the detailed technique as used {\it e.g.} by \citet[and references therein]{vdb}.
%
We use the nearest neighbour algorithm, i.e. we take $2^n$ pixels of the original image, and put the information from \textit{one of those}, the nearest neighbour, to the resulting pixel in the degraded image. This way we keep the same noise level in the degraded images. There is no correction made for seeing. Since for near-IR observations of high redshift galaxies the rest-frame wavelength is in the optical, the image degradation was made for the $B$-band images.

\begin{figure*}
\includegraphics[width=0.8\textwidth]{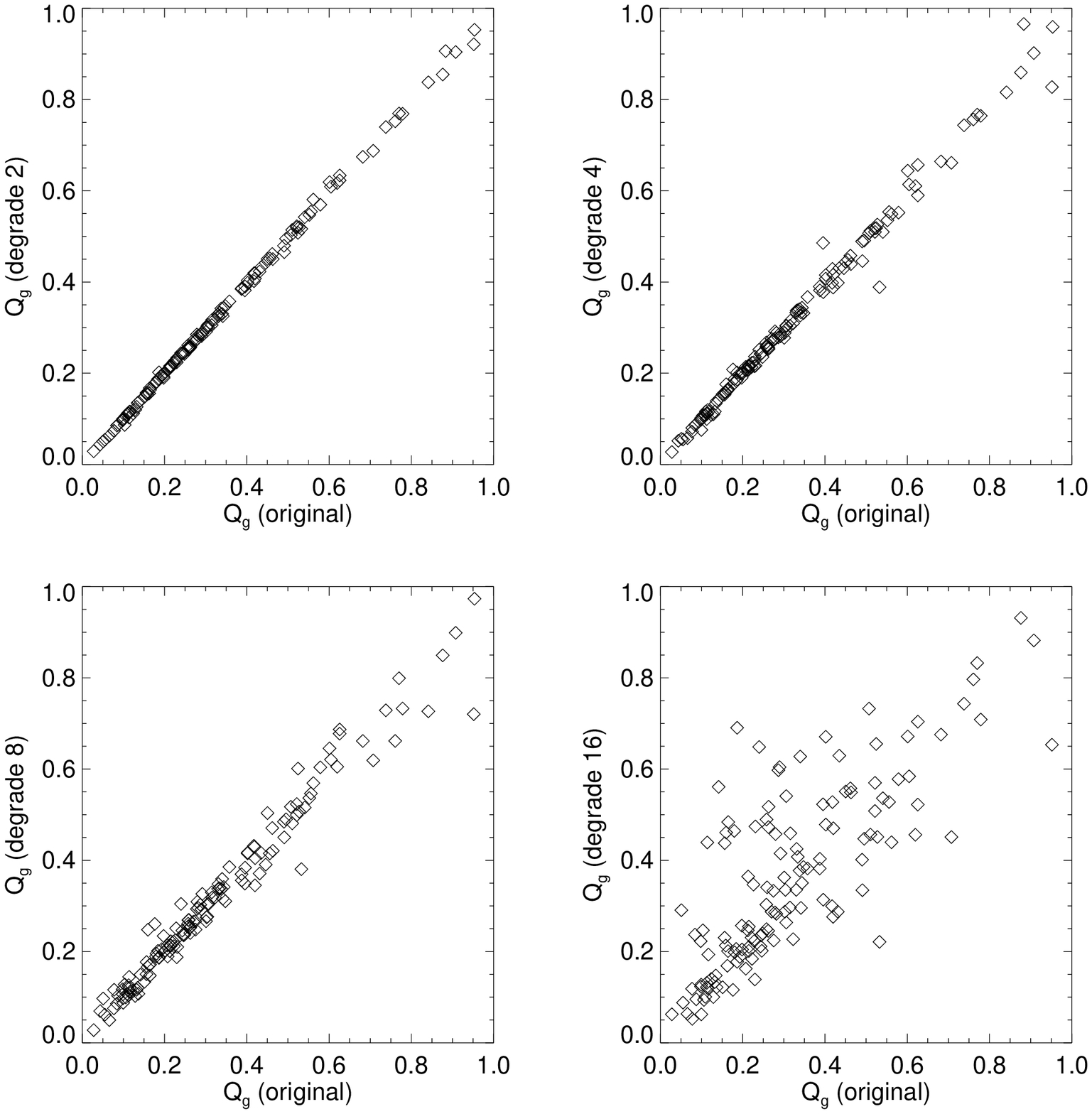}
\caption{Correlation between the $Q_g$ values after degradation by different factors and the original values. Up to a factor of 8 the correlation is very good. \label{degr1}}
\end{figure*}

\begin{figure*}
\includegraphics[width=0.7\textwidth,angle=90]{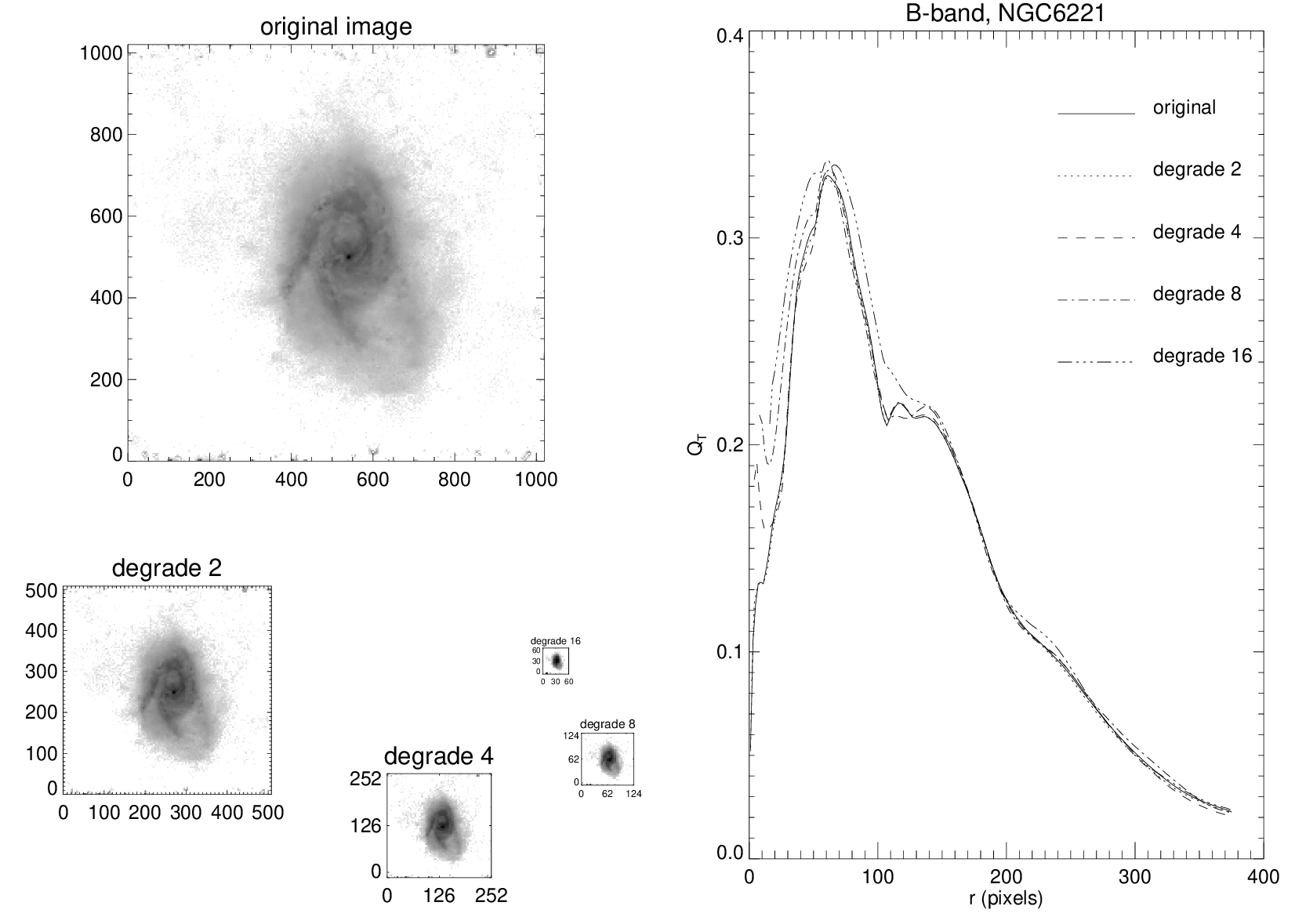}
\caption{Resulting $Q_T$ profiles for NGC 6221 after degradation are shown on the right with the radius in pixels according to the original image. On the left we show the corresponding images. \label{degr2}}
\end{figure*}

In Fig. \ref{degr1} we show the change in bar strength $Q_g$ after degrading the images by different factors. After a degradation by a factor of 2 (top left panel) or 4 (top right panel), the $Q_g$ values barely differ from the original ones, indicated by the tight correlation in $Q_g$ obtained from the original and the degraded images.  Using a factor of 8 the correlation still holds (bottom left panel), but using a factor of 16 only a small percentage of the galaxies with very low bar strengths still give nearly the same $Q_g$ as before the degradation (bottom right panel). In Fig. \ref{degr2} we show an example of an extreme case, NGC 6221, for which the whole $Q_T$-profile stayed almost unchanged even after degrading the image by a factor of 16.

\begin{figure*}
\includegraphics[width=0.6\textwidth]{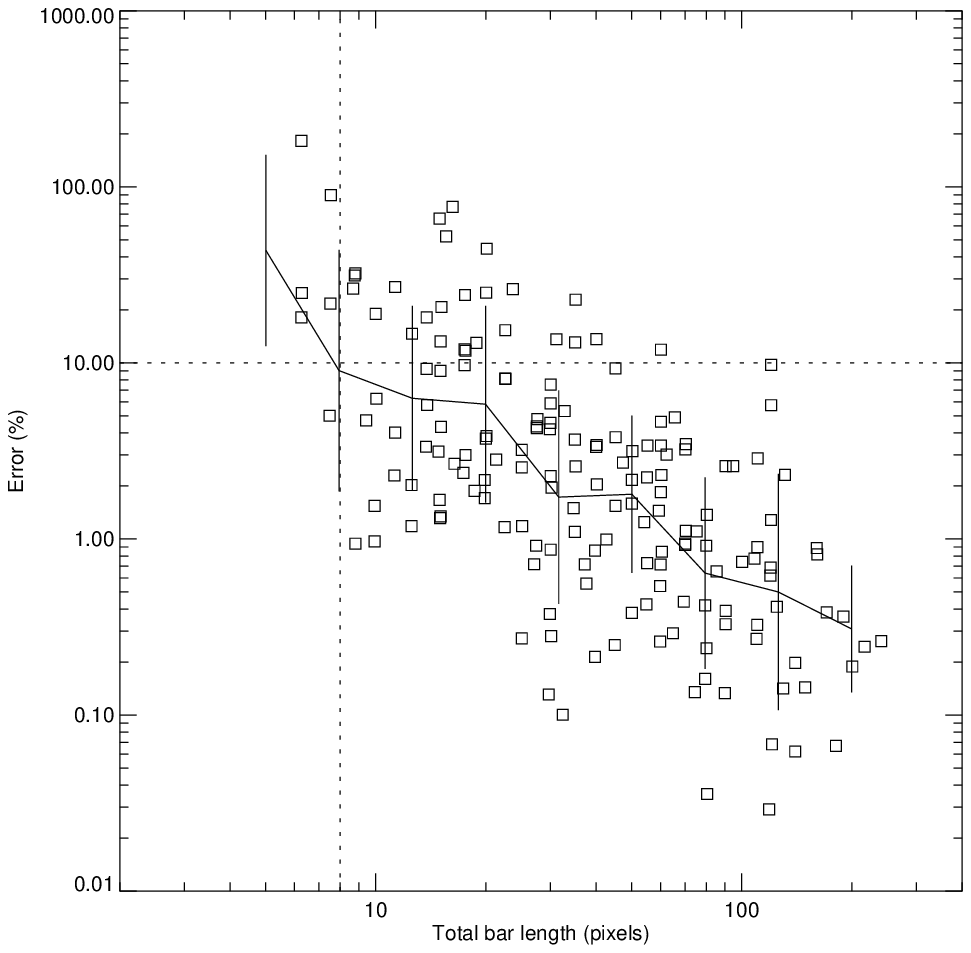} \caption{Error in per cent in relation to the total bar length in pixels. Individual galaxies are plotted with squares. Each galaxy is represented by four data points, corresponding to different degradation factors. The full line shows the average error, using logarithmic bins for the pixel size. It is clear that for bars that occupy 10 pixels or more, the average error will be below 10\% (indicated by the dashed lines).\label{degr3}}
\end{figure*}

In Fig. \ref{degr3} we translate all this into pixel sizes. For each barred galaxy in the $B$-band we calculated the bar length in pixels and the deviation from the original bar strength after the degradations by a factor of 2, 4, 8 and 16. Therefore, each of those galaxies has four corresponding points in the plot. In general, as long as the total length of the bar is at least 10 pixels, the difference between the bar strength derived from the original and the degraded image remains typically less than 10\% (indicated by the dashed lines in the figue). Thus, the $Q_g$ method can estimate bar strengths quite well, even from low resolution images. With the $0\farcs 05$ pixels from the ACS, this means that at $z=1$, for any bar that is at least $2-3$ kpc (depending on the cosmological model), we can reliably estimate its strength.

\section{Discussion}\label{discussion}
Estimating bar strengths, even in the local Universe, has proved to be far from trivial. The main parameters used to characterize bar strength are the ellipticity of the bar $\epsilon$, the bar torque $Q_g$, as well as the length and the relative luminosity of the bar. These parameters do not show similar trends in the Hubble sequence, mainly because $\epsilon$, and particularly $Q_g$ are diluted by the axisymmetric force field, largely related to the more massive bulges in the early-type galaxies \citep{lsb,laur07}. This means that massive bars ($A_2$ amplitude is large, and/or long compared to the scale length) do not necessarily have the strongest bar torques. For the characterization of bars at different redshifts all these parameters are expected to be important. This is also one of the reasons that makes our bar strength calculation particularly useful: since our method utilizes azimuthal Fourier decomposition of intensity, we obtain, besides $Q_g$, also an estimate for the relative luminosity of the bar in terms of Fourier density amplitudes. Moreover, the length of bar can be estimated based on the phase of the $m=2$ and $m=4$ Fourier amplitudes.

Questions have been raised, though, whether calculating bar strengths is feasible at high redshifts. Namely, it has been argued by \citet{marinova} that the bar torque method probably fails because (i) it is not suitable for noisy images, (ii) it is not applicable for images with low resolution, and (iii) it makes explicit use of the vertical scale height and morphological type, which can not always be reliably derived for high redshift galaxies. However, it is important to note that two types of bar strength methods have been used in the literature: the Cartesian method \citep*{quillen}, applied by \citet{bb} and \citet{block02}, and the polar method (\citealp{ls02}; \citealp[see also][]{salo99}), applied by \citet{lsb}, \citet{lsbv,l06} and \citet{bls,bvsl,buta06}, and also used in the present study. The Cartesian method is indeed very sensitive to pixel-to-pixel noise, easily leading to spurious peaks in the force ratio, particularly in the outer parts of the images where the S/N-ratio is small. However, the polar method is not very sensitive to noise, because the use of a Fourier density decomposition prior to the calculation of the forces implies averaging over the azimuthal angle (for a comparison between the polar and Cartesian methods, see \citet*{salo04}). In fact, since the calculation of bar torques integrates over the whole luminosity distribution, the force evaluation (with the polar method) can be expected to be even less sensitive to noise than the method where ellipses are fitted to the intensity contours, averaged over very narrow radial bins. The general characteristics of the Cartesian and polar methods were compared by \citet{ls02}; the robustness of the polar method with respect to noise is also strikingly illustrated in \citet{salo04}.

The two other issues mentioned by \citeauthor{marinova} have been addressed in this study, and partly also by \citet{ls02}. We found that our $Q_g$-method is not overly sensitive to image resolution: the local {\sc osubsgs} galaxies observed with the pixel size of $0\farcs 44$ can be degraded by a factor of 8, and we still obtain an indistinguishable average bar strength. It seems that the method can be applied within a 10\% accuracy, as long as the total bar length is at least about 8 pixels. As a measure of comparison, we note that \citet{jogee} find bars in their \textsc{gems} survey with a semi-major axis length of $0\farcs 15-2\farcs2$. The ACS has $0\farcs 05$ pixels, so these bars have total lengths ranging from $6-88$ pixels. Therefore we conclude that the $Q_g$-method would give a reliable estimate of the bar strength for most of those galaxies. As stated above, for a bar length of 8 pixels we find an average error of 10\%, and this decreases to a mere 1\% when the bar length is at least 40 pixels.

The estimation of the vertical scale height in the bar torque method might appear problematic if the radial scale length of the disk is not well known. Probably no detailed 2D multi-component decompositions can be applied at high redshifts.  However, a one dimensional bulge/disk decomposition using an azimuthally averaged image is even better to estimate $h_R$ in a reliable manner. It was shown by \citet{l06} that the 1D method gives very similar $h_R$-values to the more sophisticated 2D multi-component approach. In order to estimate the vertical scale height, we should also be able to classify the morphological type of the galaxy. However, in order to use the empirical relation of \citet{degrijs}, it is enough to divide the disk galaxies into 3 different Hubble type bins. Also, no \textit{ad hoc} knowledge of the barred properties is required. In case we have no information of the morphological type of the galaxy, we can use the $z_0/r_{K_{20}}$ ratio to estimate the vertical thickness, in which case we can still measure the bar strength within $\pm$ 15\%. Finally, it was shown by \citet{ls02} that the bar torque method is practically independent of the model used for the vertical distribution of the light, as long as the scale height is fixed; therefore for simplicity an exponential function has generally been used.

Based on the above discussion we believe that the bar torque approach is promising for characterizing bars even at high redshifts, providing useful complementary information to the bar ellipticity. Both the ellipticity and the bar torque are affected by the underlying axisymmetric potential, usually the bulge. However, the bar torque seems to be a better indicator of the systematic differences between bars in the Hubble sequence: $Q_g$ increases toward later types \citep{lsbv,laur07}, whereas $\epsilon$ is maintained nearly constant in all Hubble types \citep{sheth06,laur07,marinova}.

The Fourier method is also very useful for identifying bars in galaxies, as shown for the {\sc osubsgs} galaxies by \citet{lsb}. In that study several different types of diagrams were used for the identification, but the most important criteria were to detect the $m=2$ Fourier amplitude above the noise in the bar region, and to look at the phases of the $m=2$ and $m=4$ Fourier components: the phases are maintained nearly constant in the bar region.  This approach is expected to pick up classical bars, but not the more oval-like or spiral-like bars, generally classified as SAB types in RC3. The second criterion is actually the same as in the approach where ellipticities are used to identify bars: besides requiring that the ellipticity increases steadily to a global maximum, the position angle is also assumed to remain constant in the bar region. Therefore it is not unexpected that similar bar frequencies in the {\sc osubsgs} sample have been obtained by applying the Fourier method by \citet{lsb}, and by the ellipticity approach by \citet{marinova}. Using an inclination limit of 60\degrees, the found bar frequencies are 62$\%$ and 60$\%$, respectively. The ellipticity approach has been used also by \citet{sheth06}, applied for a magnitude-limited sample of 2MASS galaxies: using an inclination limit of 65\degrees they found that 59$\%$ of the galaxies are barred, which is exactly the same value as found by us using the Fourier approach with the same inclination limit \citep{lsbv}. In all these studies the number of bars increases by some 10\% if the criterion of the Fourier phases/position angles is omitted. It seems that both methods pick up the classical bars well, with results slightly larger than the 56$\%$ of SB-type bars visually identified by \citet{osu} in the near-IR for the {\sc osubsgs} sample. This is good news because it means that both Fourier and ellipticity methods can be used to estimate bar frequencies at high redshifts. However, it is worth noticing that the above comparisons have been made for spiral galaxies, which generally have small bulges. If the bulges are large, which is the case for some early-type disk galaxies, the bars might be overshadowed by the light of the bulge in the surface brightness profile. In that case, due to the azimuthal averaging, the Fourier method is expected to be more sensitive to pick up the non-axisymmetric component, even if that structure is not prominent in the contours of the two-dimensional surface brightness distribution, used to identify bars in the ellipticity approach.
 
Probably the most unexpected result from this paper is that the bar strengths in the $B$-band are larger than in the $H$-band, especially since bars are found to be more prominent in the infrared \citep[e.g.][]{osu}. We found (Section \ref{comparisons}) that different bulge dilution in optical and infrared explains the larger bar strengths in the optical. The bulge is less prominent in the optical, underestimating the radial force due to the bulge and hence artificially enhancing the bar strength.
 
\section{Summary and conclusions}\label{conclusions}
We have used a gravitational bar torque method to compare bar strengths derived from the $B$ and $H$-band images for the galaxies in the {\sc osubsgs} sample. We presented four different ways to do the comparison. Further we subjected the method to various tests simulating a high redshift environment. We estimate $z_0$ through $r_{K_{20}}$, and degrade the $B$-band images using a nearest neighbour algorithm before calculating the bar strengths. Our main conclusions can be summarized as follows:

(i) Quite unexpectedly, we found that bars appear to have stronger torques in the optical than in the near-IR: the conversion factor between $B$ and $H$-bands is 1.25. We also showed that this is due to reduced dilution of radial forces by relatively smaller bulges in the optical.

(ii) The scaled bar lengths, $barlen$/$r_{K_{20}}$, and $r_{Q_g}/barlen$ are similar in the $B$ and $H$-bands, with the bars in the $B$-band being on average slightly shorter than bars in the $H$-band.

(iii) An approximation of the vertical scale height while calculating $Q_g$ can be estimated from $r_{K_{20}}$. Resulting bar strengths will be overestimated for early types and underestimated for late types, by less than 15\%.

(iv) The $Q_g$ method is found to work well even on low resolution images. We degraded our $B$-band images using the nearest neighbour algorithm and find that as long as the total bar length is at least $\sim 10$ pixels, the resulting $Q_g$ value is typically within 10\% of the original one.

We have shown that the $Q_g$ method provides a reliable tool for measuring bar strengths at high redshifts: with the pixel resolution of $0\farcs 05$ of the ACS, bars with $\sim 2-3$ kpc can still be reliably measured at $z \sim 1.0$.

\section*{Acknowledgments}
We would like to thank the referee, Kartik Sheth, for sharing his insights and for giving instructive comments that have improved this paper. This publication makes use of data from the Two Micron All Sky Survey, which is a joint project of the University of Massachusetts and the Infrared Processing and Analysis Center/California Institute of Technology, funded by the National Aeronautics and Space Administration and the national Science Foundation. The support from the Academy of Finland is acknowledged; TSX would like to thank also CIMO and Emil Aaltonen Foundation for their financial support.


\label{lastpage}

\clearpage

\end{document}